\documentclass[prl,twocolumn,superscriptaddress,floatfix,amsmath]{revtex4-1}
\usepackage{graphicx}
\usepackage[normal]{subfigure}
\usepackage[usenames,dvipsnames]{xcolor}

\usepackage[ansinew]{inputenc}
\newcommand{\be}{\begin{equation}}
\newcommand{\ee}{\end{equation}}

\begin{document}

\title{Solid-state quantum optics with quantum dots in photonic nanostructures}

\author{Peter~Lodahl}\email{lodahl@nbi.ku.dk} \homepage{www.quantum-photonics.dk}
\author{S{\o}ren~Stobbe}
\affiliation{Niels Bohr Institute, University of Copenhagen, Blegdamsvej 17, DK-2100 Copenhagen, Denmark}

\date{\today}

\begin{abstract}
Quantum nanophotonics has become a new research frontier where quantum optics is combined with nanophotonics in order to enhance and control the interaction between strongly confined light and quantum emitters. Such progress provides a promising pathway towards quantum-information processing on an all-solid-state platform. Here we review recent progress on experiments with single quantum dots in nanophotonic structures. Embedding the quantum dots in photonic band-gap structures offers a way of controlling spontaneous emission of single photons to a degree that is determined by the local light-matter coupling strength. Introducing defects in photonic crystals implies new functionalities. For instance, efficient and strongly confined cavities can be constructed enabling cavity-quantum-electrodynamics experiments. Furthermore, the speed of light can be tailored in a photonic-crystal waveguide forming the basis for highly efficient single-photon sources where the photons are channeled into the slowly propagating mode of the waveguide. Finally, we will discuss some of the surprises that arise in solid-state implementations of quantum-optics experiments in comparison to their atomic counterparts. In particular, it will be shown that the celebrated point-dipole description of light-matter interaction can break down when quantum dots are coupled to plasmon nanostructures.
\end{abstract}

\maketitle

\section{Introduction}
Quantum optics spectroscopy with atoms has been one of the most important areas of physics in the twentieth century. For example, the extremely accurate measurements of the Lamb shift~\cite{Lamb1947} led to the theory of quantum electrodynamics. In this theory, even in the absence of any matter, a fluctuating vacuum electromagnetic field is present and it gives rise to important physical processes such as spontaneous emission of photons, the Lamb shift of atomic transitions, and Casimir forces between solid materials~\cite{Milonni}. In parallel with these developments, the inventions of the transistor and the integrated circuit have paved the way for modern computer technology based on electronics and epitaxial crystal-growth techniques has enabled accurate fabrication of semiconductor heterostructures. Combining the planar technology of integrated circuits with optics has led to a vast range of new opportunities for exploiting quantum optics in solid-state implementations. The potential benefit of such an approach is large, since integrating quantum functionalities onto a chip combining optics and electronics could lead to scalable quantum-information processing. In the present review we will discuss quantum dots embedded in nanophotonics structures as a way of locally enhancing the interaction between light and matter. We will describe the fundamental understanding of the optical properties of quantum dots in nanophotonics and discuss various applications of exploiting nanophotonics to create highly efficient and coherent single photons on demand, which is an important requirement for quantum-information processing.

\section{Self-assembled quantum dots as photon sources in nanophotonics}
\begin{figure}
\includegraphics[width=\columnwidth]{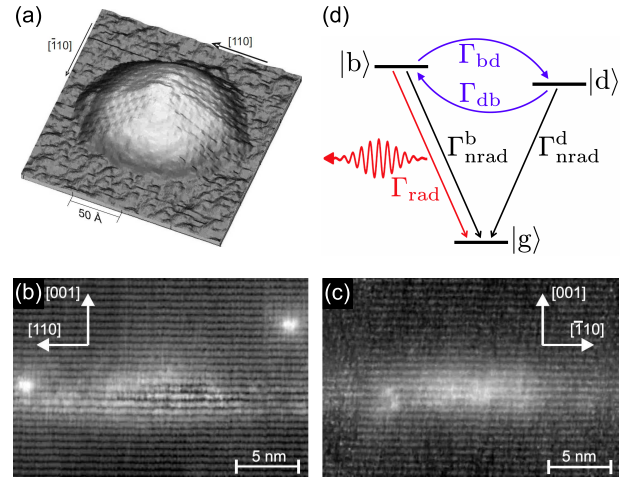}
\caption{\label{Figure_QD_overview} Structural and optical properties of quantum dots. (a)-(c) Scanning tunneling micrographs of InAs quantum dots obtained in (a) plan-view of an uncapped quantum dot and (b)-(c) cross-sectional view of a capped quantum dot~\cite{Eisele2008}. The capping leads to a complex redistribution and intermixing of indium and gallium atoms. (d) Level scheme for exciton states in a quantum dot showing the bright ($|b\rangle$) and dark ($|d\rangle$) exciton states and their decay channels to the ground state ($|g\rangle$), as discussed in the main text. (a)-(c) Reprinted with permission from Ref.~\onlinecite{Eisele2008}, copyright (2008) American Institute of Physics.}
\end{figure}
Quantum dots are solid-state quantum emitters that are made up of thousands of atoms but nonetheless have "atomic-like" optical properties. Notably, quantum dots are excellent single-photon sources with a high degree of purity~\cite{Michler2000}, as a consequence of the size-confinement effect of the trapped electrons and holes since the quantum dots are embedded in materials with a larger electronic band gap. In the present paper, we will consider only the case of self-assembled InAs/InGaAs quantum dots embedded in a GaAs barrier material, since they constitute the most well-characterized quantum-dot system with sufficiently good optical properties that quantitative quantum optics experiments are possible. For example, the single-photon wavepackets emitted from quantum dots have been shown to be coherent at low temperatures in two-photon interference experiments~\cite{Santori2002}, which is an important necessity for employing such a single-photon source in quantum-information processing. Uncapped quantum dots are essentially pyramidically shaped as shown in the scanning tunneling micrograph in Fig.~\ref{Figure_QD_overview}(a)~\cite{Eisele2008}. However, the complex intermixing between indium and gallium that occurs during growth of the capping layer leads to both highly fluctuating confinement potentials as well an overall in-plane asymmetry as shown in Figs.~\ref{Figure_QD_overview}(b)-(c); both effects have important implications for the optical properties of quantum dots as we will discuss in detail below.

The essential quantum dot level scheme in the exciton picture is shown in Fig.~\ref{Figure_QD_overview}(d), which is relevant when pumping the quantum dots weakly so that at maximum one excitation (i.e. one exciton) is created at a time in the quantum dot. The quantum-dot ground state, $|g\rangle$, corresponds to an empty conduction band and a full valence band, while the first excited state is populated by promoting an electron to the conduction band. It is convenient to implement a quasi-particle description where the excited state is populated by an exciton that corresponds to a correlated electron-hole pair while the ground state corresponds to no exciton. Due to the exchange interaction, the first excited state splits into four states: two bright exciton states, $|b\rangle$, and two dark exciton states, $|d\rangle$~\cite{Bayer2002Exchange,Poem2010}. These bright states are further split due to the lack of rotational symmetry of quantum dots and the anisotropic exchange interaction between electrons and holes. The bright excitons are optically active and can recombine by emitting photons (at a rate denoted $\Gamma_\text{rad}$) polarized linearly along the [110] and [$\bar{1}1$0] crystallographic directions of the zinc-blende crystal lattice of GaAs that are perpendicular to the [001] growth direction. In addition to radiative recombination, the bright excitons can also recombine non-radiatively (at a rate denoted $\Gamma^\text{b}_\text{nrad}$), which is likely due to trapping of carriers at defects at the interface between the quantum dot and the surrounding matrix~\cite{Stobbe2009} or in local inhomogeneous potentials formed inside the quantum dot~\cite{Stobbe2010}. The radiative decay of dark excitons is forbidden by selection rules, but non-radiative processes occur (at a rate denoted $\Gamma^\text{d}_\text{nrad}$). It has been shown experimentally that $\Gamma^\text{b}_\text{nrad}=\Gamma^\text{d}_\text{nrad}$~\cite{Stobbe2009}, which is due to the small energy difference between the bright and dark states implying that the coupling to defect states is identical. Bright and dark excitons are furthermore connected by spin-flip processes (by rates denoted $\Gamma_\text{bd}$ and $\Gamma_\text{db}$) that convert a bright exciton into a dark exciton or vice versa~\cite{Smith2005,Johansen2010}. These processes can be mediated by the short-range electron-hole exchange interaction~\cite{Roszak2007,Roszak2008Erratum} or spin-orbit interactions~\cite{Tsitsishvili2005,Liao2011} combined with acoustic phonons, or, for small bright-dark exciton splittings due to external magnetic fields, hyperfine interactions with the nuclei~\cite{Kurtze2012}. In the following we consider only experiments without an applied magnetic field. More work is needed to identify the dominant spin-flip mechanism but presently it appears that the spin-orbit coupling agrees best with experiments~\cite{Liao2011}. The bright-bright exciton spin-flip processes are found theoretically to be much slower than any of the processes discussed above~\cite{Tsitsishvili2010}, which is confirmed by measurements of the anisotropy of the radiative decay rate of bright excitons in photonic crystals~\cite{Wang2010}. This means that the two bright states are effectively decoupled, which justifies considering only one bright state as in Fig.~\ref{Figure_QD_overview}(d). The bright-dark spin-flip processes are generally slow compared to the radiative and non-radiative decay processes, but are essential to include in the analysis of the spontaneous-emission dynamics. Thus, their presence implies that the population of dark excitons that will be generated for non-resonant exciation of quantum dots will eventually lead to photon emission since the dark excitons can spin flip to bright excitons and radiatively recombine.

By solving the rate equations for the three-level exciton scheme in the case of non-resonant excitation where bright and dark excitons are populated equally, the emission from the quantum dots is predicted to decay bi-exponentially in time. Fitting the decay curves allow determining independently the rates for the three processes: radiative decay $\Gamma_{\text{rad}}$, nonradiative decay $\Gamma_{\text{nrad}},$ and spin-flip processes $\Gamma_{\text{db}}=\Gamma_{\text{bb}}$~\cite{Wang2011} where the latter equality holds at the typical temperatures applied in experiments where the probabilities for phonon absorption and emission are essentially equal. The ability to extract the radiative rate opens for important new opportunities, since it enables direct measurements of the quantum-dot oscillator strength and allows using quantum dots as sensitive probes of the local light-matter coupling strength in complex nanophotonic structures, as will be discussed in the next section. We note in passing that experiments on simple nanostructures with well-understood optical properties have proven that quantum dots in dielectric nanostructures are well described by dipole theory~\cite{Johansen2008}, which is confirmed by thorough theory~\cite{Kristensen2012}. In contrast in metallic nanostructures, where plasmons can be excited, the dipole approximation was recently found to break down~\cite{Andersen2011}, as will be discussed in detail in the end of the present review.

The oscillator strength, $f$, is a dimensionless quantity characterizing the coupling strength of an emitter to the electromagnetic field and is linked to the radiative rate in a homogeneous medium through~\cite{Siegman_laserbog}
\begin{equation}
f(\omega) = \frac{6 \pi m_0 \epsilon_0 c^3}{n q^2 \omega^2}\Gamma_\text{rad}^\text{hom}(\omega),
\end{equation}
where $m_0$ is the electron rest mass, $\epsilon_0$ is the vacuum permittivity, $c$ is the speed of light in vacuum, $n$ is the refractive index of the medium, $q$ is the electron charge, $\hbar\omega$ is the transition energy of the emitter, and $\Gamma_\text{rad}^\text{hom}$ is the radiative decay rate of the emitter in the homogeneous medium. The oscillator strength is determined by the exciton wave function and therefore depends on the confinement potentials. Two different confinement regimes exist referred to as strong and weak confinement, respectively, depending on whether the confinement potential dominates over electron-hole Coulomb attraction or vice versa. Standard-sized quantum dots, cf. Fig.~\ref{Figure_QD_overview}(b)-(c), are typically well-described in the strong-confinement regime, while large quantum dots in the weak confinement regime have been predicted to possess a giant oscillator strength~\cite{Hanamura1988,Andreani1999}. Experimental indications of a giant oscillator strength has been reported in the literature~\cite{Hours2005}. However, the importance of taking proper account of non-radiative recombination processes for large quantum dots was demonstrated in Ref.~\onlinecite{Stobbe2010}, which shows that indirect methods of determining the oscillator strength from, e.g., the vacuum Rabi splitting~\cite{Reithmaier2004,Reitzenstein2009} are not reliable probably due to collective effects of several emitters~\cite{Madsen2012NJP}. In the strong-confinement regime, an oscillator strength of up to $f=14.5$ has been reported~\cite{Johansen2008}, which is an order of magnitude larger than typical values for atomic emitters, which is a direct consequence of the many-particle nature of quantum dots.

\begin{figure}
\includegraphics[width=\columnwidth]{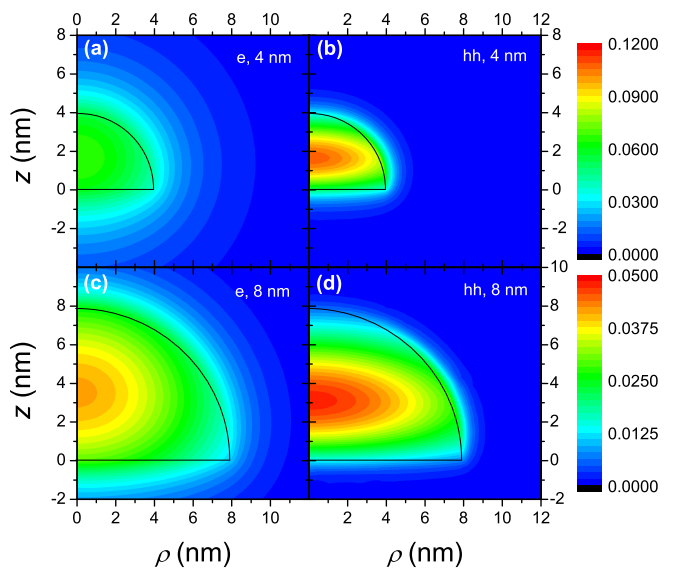}
\caption{\label{Figure_wavefunction_overlaps} The origin of the energy dependence of the oscillator strength for quantum dots. Contour plots in the radial plane $(\rho,z)$ of the amplitude of electron (e) and heavy-hole (hh) wave functions confined in axially symmetric quantum dots calculated in the envelope-function approximation. Due to the smaller effective mass of the electron, its wave function is expelled from the quantum dot when the size decreases, contrary to the heavy holes, which remain confined in the quantum dot. This effect governs the emission-energy dependence of the oscillator strength in strongly confined quantum dots. Reprinted with permission from Ref.~\onlinecite{Stobbe2009}, copyright (2009) The American Physical Society.}
\end{figure}

Having measured the oscillator strength, detailed information about the confinement of excitons in the quantum dots can be extracted. In the strong confinement regime the oscillator strength can be expressed as
\be
f(\omega) = \frac{E_\text{p}}{\hbar \omega} \left| \int \text{d}^3\mathbf{r} F^\ast_\text{e}(\mathbf{r})F_\text{h}(\mathbf{r}) \right|^2\label{eq:OS1},
\ee
where $E_\text{p}$ is the Kane energy and $F_\text{e}(\mathbf{r})$ and $F_\text{h}(\mathbf{r})$ denote the electron and hole envelope wave functions respectively. The envelope wave functions are obtained by solving an effective-mass Schrödinger equation~\cite{Efros1982}. Using this relation the measured energy dependence of the oscillator strength~\cite{Johansen2008} can be related to the dependence of the electron-hole wave functions on quantum-dot size as illustrated in Fig.~\ref{Figure_wavefunction_overlaps}. Due to confinement and strain, the valence-band states are dominated by the heavy holes, which have a much larger effective mass than that of the electrons. As a consequence, for decreasing quantum-dot size (increasing emission energy) the electron wave function is expelled from the confining quantum dot potential more than the hole wave functions~\cite{Stobbe2011}. This leads to a decrease in oscillator strength with increasing emission energy, which has been confirmed by experiments and detailed numerical modeling~\cite{Stobbe2009}.

\section{Spontaneous emission control in photonic crystals}

The potential of periodic dielectric structures for manipulating light was proposed several decades ago~\cite{Bykov1975,Yablonovitch1987PC,John1987} and was inspired by the physics of electron scattering on atomic crystal lattices. Since then, the research field of photonic crystals has been blossoming enabling a multitude of novel photonics functionalities that can be integrated on an optical chip, including slow-light waveguides, nanolasers, and tunable filters~\cite{JoannopoulosBook}. More recently, the quantum-optics community has started to show a strong interest in photonic crystals due to their ability to strongly enhance the interaction between light and matter providing a possible pathway to scalable all-solid-state quantum-information processing~\cite{Shields2007,OBrien2009}. A paradigmatic quantum-optics setting is that of a single quantum emitter radiating a single-photon wavepacket. Embedding the quantum emitter into a photonic-crystal heterostructure offers a way of controlling the spontaneous photon emission whereby the emission rate can be either suppressed or enhanced and the photon can be channeled into predetermined optical modes. This ability to control spontaneous emission is beneficial for highly efficient single-photon sources for quantum-information applications, but also of much wider use, e.g., for efficient light-emitting diodes (LEDs) or energy harvesting~\cite{Bermel2007}.

\begin{figure}
\includegraphics[width=\columnwidth]{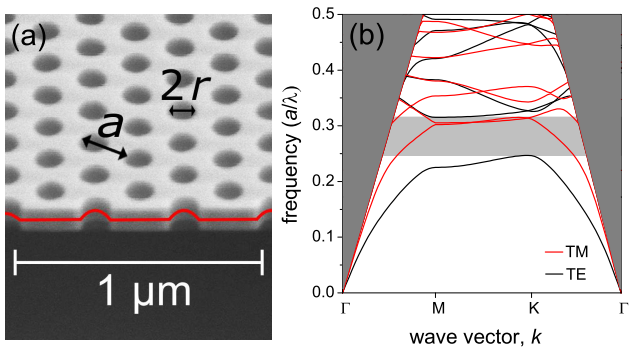}
\caption{\label{Figure_PC_a_240nm_r_72nm_h_150nm} (a) Scanning electron micrograph of a photonic-crystal membrane with lattice constant $a$ and hole diameter $2r$. In QED experiments a layer of quantum dots is embedded in the center of the membrane, which is illustrated by the line. (b) Calculated photonic band structure for a GaAs photonic-crystal membrane where $a/\lambda$ is the relevant scaled frequency with $\lambda$ the wavelength in vacuum. On the horizontal axis is shown the in-plane component of the wave vector along the relevant symmetry directions of the triangular photonic lattice. Above the light line (dark gray) the is no optical confinement. In the 2D photonic band gap (light gray) no TE-modes exist below the light line.}
\end{figure}

Experimental work on spontaneous emission in photonic crystals has employed various type of quantum emitters including dye molecules~\cite{Petrov1998,Megens1999,Koenderink2002,Nikolaev2008}, quantum wells~\cite{Fujita2005}, nitrogen vacancies~\cite{Faraon2012}, colloidal quantum dots~\cite{Lodahl2004,Barth2006,Nikolaev2007,Leistikow2011}, and self-assembled quantum dots~\cite{Kress2005PRB,Kaniber2007,Noda2007,Julsgaard2008,Wang2011} in either 2D or 3D photonic crystals. In the current review, we will focus on the research on self-assembled quantum dots in 2D photonic-crystal nanomembranes that is presently the most mature platform for quantum-optics experiments. Indeed, the ability to address single quantum dots with excellent and well-characterized optical properties ensures the viability of this approach. Figure~\ref{Figure_PC_a_240nm_r_72nm_h_150nm}(a) shows an example of a 2D photonic crystal membrane fabricated in GaAs that contains a single layer of InGaAs quantum dots in the membrane center. The large refractive index of GaAs (around $n=3.5$ for the typical wavelengths and temperatures in the discussed experiments~\cite{Gehrsitz2000}) ensures that a thin photonic-crystal membrane very efficiently confines light since the 2D photonic band gap can suppress radiation in the plane of the membrane while total internal reflection strongly suppresses light leaking vertically out of the structure. Figure~\ref{Figure_PC_a_240nm_r_72nm_h_150nm} shows the band structure calculated for a GaAs photonic-crystal membrane with lattice constant $a=240\:\text{nm}$, hole radius $r=72\:\text{nm}$, and membrane thickness $h=150\:\text{nm}$. For the transverse-electric (TE) modes, a frequency region is observed where no modes exist below the light line, which is the 2D band gap. Above the light line the modes are not confined by total-internal-reflection to the membrane and are leaky. For transverse-magnetic (TM) modes no band gap appears. However, since the self-assembled quantum dots have their dipole orientation in the plane of the photonic crystal due to the predominantly heavy-hole character of the excitons, very strong suppression of spontaneous emission is possible in the membranes.

Two different types of experiments are typically performed by recording either emission spectra or time-resolved decay curves. While both approaches probe spontaneous emission, the former measurements depend on both the emission and the propagation of light in the photonic crystals, thus rendering a quantitative account very challenging. Indeed the propagation of the emitted photon from the quantum dot to the detector is a rather complex process that depends sensitively on microscopic details such as the exact positions of the quantum dot and the detector, intrinsic fabrication imperfections in the vicinity of the quantum dot, and interference between different propagation paths for the photon traveling from the quantum dot to the detector~\cite{Madsen2012NJP}. Such detailed knowledge is not available in present experiments and consequently modeling of quantum dot emission spectra in photonic-crystal cavities have relied on multi-parameter fitting of theory to experiment~\cite{Laucht2009PRL}. In contrast, time-resolved experiments probe the intrinsic rate of spontaneous emission of a quantum dot, which can be directly obtained since the time delay associated with propagation of the photon from the emission to the detector is usually much shorter, and hence negligible, compared to the lifetime of the decay. As a consequence, time-resolved measurements can be used as sensitive probes of the local environment of the quantum dot, including both the local light-matter interaction strength~\cite{Wang2011} and the effective phonon density of states responsible for phonon decoherence~\cite{Madsen2012}.

The essential quantity describing light-matter interaction in a nanophotonic environment is the local density of optical states (LDOS) projected along the orientation of the transition dipole moment of the emitter in consideration~\cite{Sprik2006}. The LDOS can be expressed as~\cite{Vats2002,NanoOpticsBook2012,Stobbe2009}
\be
\rho(\mathbf{r},\omega,\mathbf{\hat{e}}_\mathbf{d}) =
\frac{1}{V}\sum_{\mathbf{{k}}} |\mathbf{\hat{e}}_\mathbf{d} \cdot
\mathbf{\hat{e}}_\mathbf{{k}}|^2
\left|f_\mathbf{{k}}(\mathbf{r}) \right|^2
\delta(\omega-\omega_\mathbf{{k}}),\label{eq:LDOS_def}
\ee
where $\mathbf{\hat{e}}_\mathbf{d}$ and $\mathbf{\hat{e}}_\mathbf{k}$ are unit vectors specifying
the direction of the quantum dot transition dipole moment and the electromagnetic wavevector, respectively, $f_\mathbf{{k}}(\mathbf{r})$ denotes the spatial profile of the mode functions that the electromagnetic field is decomposed into, and $\delta(\omega-\omega_\mathbf{{k}})$ is a Dirac delta function that peaks at frequencies $\omega$ matching any of the eigenfrequencies $\omega_\mathbf{{k}}$ of the photonic structure. Finally ${\mathbf{r}}$ is the position of the dipole emitter. It is often convenient to express the LDOS in terms of the electric-field Green's tensor, $\mathbf{G}(\mathbf{r},\mathbf{r}',\omega)$, as~\cite{Vats2002,NanoOpticsBook2012,Stobbe2009}
\begin{equation}
\rho(\mathbf{r},\omega,\mathbf{\hat{e}}_\mathbf{d}) = \frac{2 \omega}{\pi c^2} \left( \hat{\mathbf{e}}_\mathbf{d}^T \cdot \mathrm{Im}\left\{\mathbf{G}( \mathbf{r},\mathbf{r},\omega ) \right\} \cdot \hat{\mathbf{e}}_\mathbf{d} \right),\label{eq:LDOS_in_terms_of_G}
\end{equation}
where $\mathrm{Im}\left\{ \ldots \right\}$ denotes the imaginary part.
Since the Green's tensor is a propagator of the electric field it is observed that the LDOS is describing the self-interference of the emitter.

In the weak-coupling regime of light-matter interaction, the Wigner-Weisskopf approximation can be invoked in the description of spontaneous-emission dynamics. The approximation is applicable when the frequency variation of the LDOS is modest over the spectral linewidth of the emitter, which is often an excellent approximation apart from the case of narrow band cavities where phenomena like strong coupling~\cite{Yoshie2004,Hennessy2007} and non-Markovian dynamics~\cite{Madsen2011,Majumdar2012} are beyond the validity of the Wigner-Weisskopf approximation. Furthermore, it has
been predicted that at the edge of a photonic band gap exotic dynamics may occur such as a fractional decay where an initially excited emitter relaxes to a partly decayed steady state~\cite{John1994,Vats2002}. Such extreme non-Markovian dynamics has not yet been observed experimentally although detailed calculations indicate that they may be within reach under present experimental conditions~\cite{Kristensen2008}. Within the Wigner-Weisskopf approximation the radiative decay rate of a quantum emitter is directly proportional to the projected LDOS~\cite{Stobbe2012}, i.e.,
\be
\Gamma_{\text{rad}}(\mathbf{r},\omega,\mathbf{\hat{e}}_\mathbf{d})
=\frac{\pi q^2}{2 m_0 \epsilon_0} f(\omega)
\rho(\mathbf{r},\omega,\mathbf{\hat{e}}_\mathbf{d}).\label{eq:Gamma_rad}
\ee
This relation is very useful for a number of reasons. It directly illustrates that spontaneous-emission dynamics can be controlled by changing the LDOS, which is the essential idea of photonic crystals. Furthermore, the ability to reliably extract $\Gamma_{\text{rad}}$ provides a way of mapping the LDOS of complex nanophotonic structures such as photonic crystals, which will be presented below. Measuring the radiative decay rate is not a trivial task in solid-state systems, since non-radiative recombination processes in general are not negligible, as was discussed above. We note that such intrinsic non-radiative effects are independent of the LDOS, which is fundamentally different from extrinsic non-radiative coupling that may occur, e.g., due to absorption in metals. Extrinsic non-radiative effects can be taken fully into account in the LDOS formalism but they are not important in dielectric structures and thus neglected in the following.

It has been an open question in the field of photonic crystals to determine how much the radiative lifetime can be altered in order to establish the fundamental limits. In an ideal 3D photonic crystal with a sufficiently high refractive-index contrast, a photonic band gap may open and the radiative decay rate ideally vanishes meaning that an excited emitter would be unable to radiate. In reality, the observed inhibition of spontaneous emission is limited by effects such as the finite size of the samples and fabrication imperfections, and so far suppression up to a factor of $10$ has been reported using an ensemble of emitters in 3D inverse woodpile structures~\cite{Leistikow2011}. In 2D photonic crystal membranes, light leakage from coupling to radiation modes poses an upper bound on the possible suppression of spontaneous emission, but simulations of the LDOS have indicated that very large suppression of up to almost $10^2$  should be achievable in such structures~\cite{Koenderink2006}.

\begin{figure}
\includegraphics[width=\columnwidth]{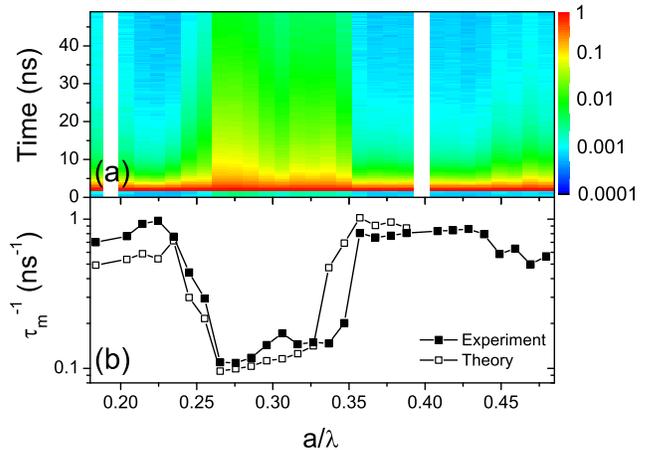}
\caption{\label{Figure_Julsgaard} Suppression of radiative decay rates observed in 2D GaAs photonic-crystal membranes. (a) Raw data of time-resolved spontaneous emission intensity for an ensemble of quantum dots and different dimensionless frequencies, $a/\lambda$. A pronounced prolongation of the decay time of the ensemble is observed in the frequency range of the photonic band gap. (b) Inverse mean lifetime, $\tau^{-1}$, extracted from the data in (a) and compared to the inverse mean lifetime expected from theory. (a) Reprinted with permission from Ref.~\onlinecite{Julsgaard2008}, copyright (2008) American Institute of Physics. (b) Reprinted with permission from Ref.~\onlinecite{Hvam2011}, copyright (2011) John Wiley and Sons.}
\end{figure}

LDOS effects in photonic crystals may be probed in time-resolved spectroscopy using either single emitters or ensembles of emitters. The former corresponds to the setting of single-photon sources for quantum-information processing while the latter would be the regime of complex devices like LEDs and lasers. Also conceptually, the information extracted from the two different situations differs. With ensembles, the overall decay of the emitters is studied while it is not possible to extract the individual decay rates of emitters from the generally highly multi-exponential decay curves~\cite{vanDriel2007}. Figure~\ref{Figure_Julsgaard} shows measurements of the decay dynamics of an InAs quantum dot ensemble in 2D photonic band gap structures and pronounced modifications are observed that can be understood qualitatively by theory~\cite{Koenderink2006} after averaging the calculated LDOS in space and accounting for quantum dot fine structure~\cite{Julsgaard2008}.

\begin{figure}
\includegraphics[width=\columnwidth]{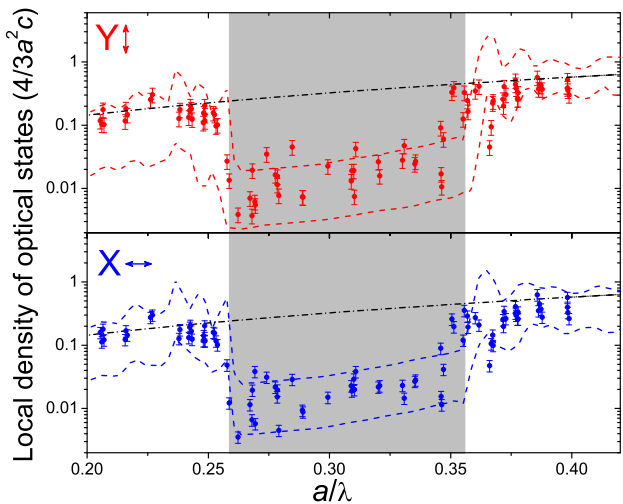}
\caption{\label{Figure_Wang} Mapping the frequency dependence of the LDOS with quantum dots for the two orthogonally polarized exciton states, $Y$ (red) and $X$ (blue). A strong suppression of the decay rates is observed within the band gap (shaded gray region) relative to the homogeneous-medium decay rate (dash-dotted curves) but significant fluctuations are observed due to the strongly varying LDOS for the different positions inside the photonic crystal at which the quantum dots are positioned. Excellent agreement with the upper and lower bounds of the calculated LDOS (dashed curves) is observed. Reprinted with permission from Ref.~\onlinecite{Wang2011}, copyright (2011) American Physical Society.}
\end{figure}

An actual mapping of the LDOS requires single-emitter experiments. By taking advantage of the detailed knowledge of the optical properties of quantum dots discussed in the previous section, it is possible to use a single quantum dot as a local probe of the LDOS and thus experimentally access the ultimate potential of photonic crystals for controlling light-matter interaction. In such experiments, spontaneous emission is recorded from single InGaAs quantum dots embedded in a GaAs photonic cystal membrane by spectral selection of a single quantum-dot line. Two different projections of the LDOS can be recorded by selecting the polarization of the emission, which corresponds to probing the two perpendicularly oriented bright exciton states. A high degree of asymmetry between the radiative decay of the two bright exciton states is generally observed~\cite{Wang2010}, which is a measure of the anisotropic vacuum fluctuations present in photonic crystals~\cite{Vos2009}. Having determined the radiative rate of single quantum dots in the photonic crystal, $\Gamma_{\text{rad}}({\bf r},\omega,{\bf \hat{e}_d})$, and in a homogenous medium, $\Gamma_{\text{rad}}^{\text{hom}}(\omega)$, the projected LDOS evaluated at the position of the emitter and at the emission frequency of the emitter can be straightforwardly obtained from
\be
\rho(\mathbf{r},\omega,\mathbf{\hat{e}}_\mathbf{d}) = \rho(\omega) \frac{\Gamma_{\text{rad}}({\bf r},\omega,{\bf \hat{e}_d})}{\Gamma_{\text{rad}}^{\text{hom}}(\omega)},
\ee
where $\rho(\omega) = \frac{n\omega^2}{3\pi^2c^3}$ is the total density of states of a homogeneous medium. The result of such an experiment was reported in Ref.~\onlinecite{Wang2011} and is shown in Fig.~\ref{Figure_Wang}. As opposed to the ensemble measurements shown in Fig.~\ref{Figure_Julsgaard}, the LDOS is here probed directly and in particular the strong fluctuations of the LDOS are directly observed, in agreement with theory. Notably, a 70-fold reduction of the radiative decay rate has been observed in photonic-crystal membranes, which shows the potential of 2D photonic crystals for controlling spontaneous emission. For such analyses, it is imperative to take the internal decay dynamics including spin-flip and non-radiative processes fully into account.

\section{Quantum electrodynamics in photonic-crystal nanocavities and waveguides}
Novel functionalities are possible when deliberately introducing defects in a photonic-crystal lattice. One popular choice is photonic-crystal nanocavities that can be applied for strongly enhancing light-matter interaction applicable for, e.g., nanolasers~\cite{Strauf2011} or cavity QED~\cite{Yoshie2004}. Two different approaches have been taken towards photonic-crystal nanocavities employing either engineered geometries~\cite{Akahane2003,Song2005} or the spontaneous formation of Anderson-localized cavities in disordered photonic-crystal waveguides~\cite{Sapienza2010,Thyrrestrup2012}. Photonic-crystal cavities distinguish themselves from other dielectric cavities by combining a high quality factor with an ultimately small mode volume~\cite{Vahala2003}, which make them particularly promising for cavity-QED experiments on coupling a single quantum emitter to a cavity mode. Depending on the magnitude of the light-matter coupling strength relative to the rates for dissipation and decoherence processes, the QED system can either be in the weak or the strong-coupling regime. In the weak-coupling regime, the rate of single-photon emission from the quantum dot is enhanced by the increased LDOS of the cavity mode, which is the Purcell effect~\cite{Purcell1946}. The rate enhancement is proportional to the ratio of cavity $Q$-factor to the mode volume, $V$, and provides a way of collecting single photons with high efficiency. It is quantified by the Purcell factor, $F_{\text{P}}$, which gauges the rate of emission into the cavity mode, $\Gamma_{\text{cav}}$, for a dipole emitter relative to the radiative spontaneous-emission rate of the same emitter in a homogeneous medium, $\Gamma_{\text{rad}}^{\text{hom}}$, with refractive index $n$. Within the approximation that the quantum emitter couples primarily to the single quasi-mode of the cavity described by the normalized spatial mode function ${\bf f}({\bf r})$, the Purcell factor is given by~\cite{GerardReview2003}
\be
F_{\text{P}}({\bf r})=\frac{\Gamma_{\text{cav}}}{\Gamma_{\text{rad}}^{\text{hom}}}= \frac{3}{4 \pi^2} \frac{Q (\lambda/n)^3}{V} |\mathbf{\hat{e}}_\mathbf{d} \cdot
\mathbf{{f}}(\mathbf{{r}})|^2 \frac{1}{1 + 4 \Delta^2 Q^2/\omega_\text{c}^2},
\label{purcell-factor-cav}
\ee
where $\Delta = \omega_e - \omega_c$ is the detuning of the quantum emitter frequency, $\omega_\text{e}$, relative to the cavity resonance frequency, $\omega_\text{c}$. A large Purcell factor requires not only a low-loss cavity (i.e., high $Q$) with small mode volume but also that the emitter spatially and spectrally matches the cavity mode. Thus, the emitter must be positioned spatially in the cavity where the electric field strength is large and with the transition dipole moment oriented along the local electric field while simultaneously being at resonance with the cavity.

The first experimental demonstrations of Purcell enhancement in photonic-crystal cavities were reported in Refs.~\onlinecite{Englund2005,Kress2005PRB} by observing the enhanced emission rate when the quantum dots were resonant with the cavity mode. The potential application of the weak-coupling regime includes an efficient single-photon source where the channeling of photons into the cavity is enhanced by the Purcell effect that also may help in overcoming dephasing processes enabling coherent single photons~\cite{Laurent2005}. The strong-coupling regime has been observed in the spectral domain by observing the avoided crossing of a single quantum-dot line when tuned into resonance with the cavity mode~\cite{Yoshie2004,Hennessy2007}. However, quantitative comparisons between experiment and theory~\cite{Madsen2012NJP} has revealed that the observation of the avoided crossing for a single quantum dot line is not necessarily a proof of vacuum Rabi splitting since additional quantum-dot lines or multiexciton transitions~\cite{Winger2009PRL} may feed the cavity, thus potentially giving rise to a collective enhancement of the Rabi splitting. In contrast, the dynamics of single-exciton transitions has proven to be well suited for extracting reliable information about the coupling strength of a single quantum dot to the cavity and quantitative agreement between experiment and theory has been found~\cite{Madsen2012NJP,Madsen2012}. Strongly-coupled quantum-dot-cavity systems have also been employed to observe the nonlinear photon-blockade effect both in continuous-wave and pulsed experiments~\cite{Englund2007,Reinhard2012}, which may enable constructing photonic gates. The strong coupling between a quantum dot and a cavity constitutes a way of entangling light and matter and various protocols based on cavity QED exist for quantum-information processing based on a coupled array of cavity QED systems~\cite{Kimble2008QuantumInternet}.

\begin{figure}
\includegraphics[width=\columnwidth]{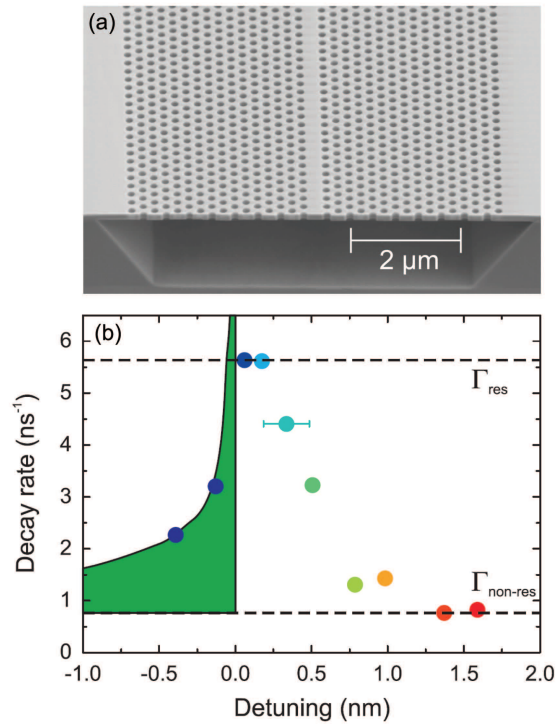}
\caption{\label{Figure_WG_beta_factor} Highly efficient single-photon source based on a single quantum dot coupled to a photonic-crystal waveguide. (a) Scanning electron micrograph of a photonic-crystal waveguide. (b) Decay rates of a single quantum dot (colored dots) temperature tuned through the slow-light region of a photonic-crystal waveguide. By comparing the decay rate on resonance, $\Gamma_\text{res}$, to the non-resonant decay rate, $\Gamma_\text{non-res}$, the $\beta$-factor can be extracted. (a) Reprinted with permission from Ref.~\onlinecite{Sapienza2010}, copyright (2010) AAAS. (b) Reprinted with permission from Ref.~\onlinecite{Thyrrestrup2010}, copyright (2010) American Institute of Physics.}
\end{figure}
Photonic-crystal waveguides provide an alternative to cavities for QED experiments. An example of a photonic-crystal waveguide implemented in a membrane of GaAs is displayed in Fig.~\ref{Figure_WG_beta_factor}(a). It was proposed that photonic-crystal waveguides do give rise to a Purcell effect since the strong dispersion of light in the waveguide enables slow light~\cite{Hughes2004,MangaRao2007PRL,LeCamp2007} in turn implying that the density of states of the waveguide mode is enhanced. In a cavity, the Purcell effect is essentially limited by the bandwidth of the cavity (cf. Eq.~(\ref{purcell-factor-cav})), although the effect of coupling to phonons in fact can broaden this bandwidth, as will be discussed in the following section. Photonic-crystal waveguides, on the other hand, offer Purcell enhancement over a much broader bandwidth. The fundamental difference compared to a cavity is that the rate enhancement in the waveguide is mediated by slow light. In a photonic-crystal waveguide, the LDOS associated with the propagating waveguide mode can be strongly enhanced and scales inversely proportional to the group velocity, $v_\text{g}$, of the guided mode. The Purcell factor in a photonic-crystal waveguide can be expressed as~\cite{MangaRao2007PRB}
\be
F_{\text{P}}({\bf r})= \frac{\Gamma_\text{wg}}{\Gamma_\text{rad}^\text{hom}} = \frac{3 \pi c^3 a}{\omega_e^2 n v_\text{g}} |\mathbf{\hat{e}}_\mathbf{d} \cdot
\mathbf{{f}}(\mathbf{{r}})|^2,
\label{purcell-factor-wav}
\ee
where $\Gamma_\text{wg}$ is the rate of channeling photons into the waveguide, $a$ is the lattice constant of the photonic crystal, $n$ is the refractive index, and $\mathbf{{f}}(\mathbf{{r}})$ is the spatial profile of the mode propagating along the waveguide. We note that the bandwidth of the photonic-crystal waveguide is determined by the dispersion of the group velocity that can be engineered by proper design of the waveguide~\cite{Frandsen2005}. Importantly, the bandwidth of Purcell enhancement can be much larger than in the case of cavities. Thus, in a cavity a high $Q$, it is required to have a large Purcell effect (cf. Eq.~(\ref{purcell-factor-cav})), but that simultaneously narrows the achievable bandwidth. Another advantage of a waveguide compared to a cavity is that in the former case single photons can be coupled directly to a propagating mode as opposed to being trapped in a localized mode. This is advantageous for applications of highly-efficient single-photon sources for quantum-information processing since the  photons collected into a waveguide could be directly usable. In contrast, in the cavity the collected photons should be coupled out from a localized nanocavity in order to be processed, which would limit the overall efficiency.

Early experiments on spontaneous emission in photonic-crystal waveguides observed a rather moderate enhancement of the emission rate of about 15\%~\cite{Viasnoff-Schwoob2005}, but this was performed on non-optimized waveguide structures. More recent experiments have focused on single quantum dots in W1 waveguides that are obtained by leaving out a single row of holes in 2D photonic-crystal membranes, which has proven to be a platform where group-velocity slow-down factors of several hundreds can be achieved~\cite{Vlasov2005}. In W1 photonic-crystal-waveguide membranes, Purcell factors of up to $5.2$ have so far been observed, cf. Fig.~\ref{Figure_WG_beta_factor}(b), in time-resolved measurements of the light leaking vertically out of the waveguide~\cite{Lund-Hansen2008,Thyrrestrup2010}, which has been backed up by experiments coupling photons out from a cleaved sample \cite{Dewhurst2010}. Furthermore the single-photon purity \cite{Schwagmann2011,Laucht2012} and rate enhancement due to Fabry-Perot resonances~\cite{Wasley2012,Hoang2012} have been studied. The magnitude of the achievable Purcell enhancement is ultimately limited by multiple scattering leading to Anderson localization in the photonic-crystal waveguide~\cite{Sapienza2010}. Importantly, the formation of Anderson-localized modes can be effectively suppressed by making the photonic-crystal waveguide shorter than the localization length, which is the average distance between scattering events. Deep in the slow-light regime the localization length can be below $10\:\mu\text{m}$~\cite{Smolka2011}, but importantly even for such short the waveguide LDOS can build up very efficiently enabling a pronounced Purcell enhancement~\cite{Sapienza2010}.

The observed Purcell factor in photonic-crystal waveguides is not yet at the level of photonic-crystal cavities. Having a large Purcell factor may be advantageous for creating indistinguishable single photons from a non-resonantly excited quantum dot where the speedup of the emission helps overcoming detrimental dephasing processes~\cite{Santori2002} or non-radiative recombination~\cite{Johansen2008}. For a range of other applications, however, such as efficient single-photon sources, nanolasers, and photon-blockade nonlinearities, it is rather the $\beta$-factor than the Purcell factor that is the relevant figure-of-merit. It is defined as
\be
\beta = \frac{\Gamma_\text{wg}}{\Gamma_\text{wg}+\Gamma_\text{rad}+\Gamma_\text{nrad}}
\ee
where  $\Gamma_\text{rad}$ is the rate of coupling to radiation modes that leak out of the photonic-crystal membrane and $\Gamma_\text{nrad}$ is the rate of intrinsic non-radiative recombination inside the quantum dot. The $\beta$-factor can be very large in photonic-crystal waveguides due to the combination of two effects: i) the rate of coupling into the waveguide, $\Gamma_\text{wg}$, is large due to the enhanced LDOS mediated by slow light and ii) the coupling to radiation modes is strongly suppressed in a photonic-crystal membrane due to 2D photonic band-gap effects, as was discussed in the previous section. In contrast, alternative methods proposed in the literature for strong interaction between a single propagating mode and a quantum emitter manipulate just one of the two processes: In plasmon nanostructures the anticipated large Purcell enhancement of the spontaneous emission into propagating plasmons relies on the slowdown of plasmons and narrow confinement~\cite{Chang2006}, i.e., mechanism i). A thorough analysis for experimentally realistic parameters and various plasmonic waveguide geometries has revealed the limited potential of this approach for quantum dots embedded in GaAs due to the large propagation loss of plasmons and the quenching of emission found when placing the quantum dots close to metals~\cite{Chen2010PRB,Chen2010OE}. Another promising approach utilizes dielectric photonic nanowires that can successfully suppress coupling to radiation modes~\cite{Claudon2010,Bleuse2011}, which is mechanism ii). However, in the nanowires the overall decay is slow implying that dephasing and non-radiative processes will limit the coherence of the single photons and the efficiency, respectively. Photonic-crystal waveguides potentially overcome these limitations and appear as a highly promising platform for implementing a single quantum dot as a giant nonlinearity capable of operating at the few-photon level~\cite{Chang2007} or for scalable quantum-information processing with deterministic single-photon sources~\cite{OBrien2009}.

A reliable measurement of the $\beta$-factor requires direct estimates of the different decay channels of a single quantum dot in the photonic-crystal waveguide. This can be achieved by time-resolved spontaneous-emission measurements. Figure \ref{Figure_WG_beta_factor}(b) shows the decay rate of a single quantum dot that is spatially matched to the waveguide and spectrally tuned by varying the temperature between 10 and $60\:K$. By increasing the temperature the quantum dot shifts towards longer wavelengths and in this process spectrally tunes from coupling efficiently to the waveguide and beyond the cutoff of the waveguide where the coupling to the waveguide ceases. The most efficient coupling is observed at the waveguide cutoff where the total decay rate $\Gamma_\text{res}=\Gamma_\text{wg}+\Gamma_\text{rad}+\Gamma_\text{nrad}$ is strongly enhanced due to the Purcell enhancement increasing $\Gamma_\text{wg}$. Above cutoff, the coupling to the waveguide is essentially turned off and non-resonant coupling to radiation modes and intrinsic non-radiative recombination is recorded: $\Gamma_\text{non-res}=\Gamma_\text{rad}+\Gamma_\text{nrad}.$ From $\Gamma_\text{res}$ and $\Gamma_\text{non-res}$ the $\beta$-factor is readily determined, and amounts to $85.4 \%$ for this particular example. From a statistical analysis of the decay rates of a number of quantum dots, an average $\beta$-factor approaching $90\%$ is inferred together with efficient coupling observed in a very large bandwidth of $20 \: \text{nm}$~\cite{Lund-Hansen2008}. The fundamental limits on the achievable $\beta$-factor have not been established yet and the reported numbers are conservative estimates. Thus, at the elevated temperatures employed for recording the non-resonant decay rate in Fig.~\ref{Figure_WG_beta_factor}(b), enhanced non-radiative processes are likely to be present and even for the detuning of $1.5\: \text{nm}$ above the waveguide cutoff, residual coupling to the waveguide could be present due to disorder-induced broadening of the band edge or phonon-assisted recombination processes. We anticipate that the achievable $\beta$-factor could ultimately be limited by the finite intrinsic non-radiative rate of the quantum dot. Employing the numbers extracted in previous work~\cite{Wang2011}, $\beta$-factors exceeding $99\%$ should be experimentally achievable, thereby illustrating the very promising potential of photonic-crystal waveguides for on chip quantum-information processing.

\section{Quantum optics with mesoscopic emitters}
The quantum theory of the interaction between light and matter, quantum electrodynamics (QED), has to a large extent been developed in the context of atomic physics. While atoms and quantum dots share a number of similarities regarding their optical properties, the analogy has its limitations. One added complexity in solid-state systems is to account for the interaction with the environment that the quantum dots are embedded into. Environmental decoherence is inevitable and in particular phonon dephasing associated with lattice vibrations of the surrounding material is important for quantum dots. This gives rise to interesting new phenomena in solid-state cavity-QED experiments including notably the observation that the effective bandwidth of a quantum dot coupled to a photonic-crystal nanocavity is much wider than what would be expected from standard QED models. This phonon-mediated Purcell effect is discussed in the first subsection below. Another concern for quantum dots in photonic nanostructures is the validity of the dipole approximation, which corresponds to treating the quantum emitter as a point source. The extent of the electron wave function for atoms is typically sub-nanometer and thus much smaller than the optical wavelength and the dipole approximation is usually an excellent approximation. However, quantum dots are mesoscopic emitters with typical lateral dimensions in the $10-100 \: \text{nm}$ range, which is not in general negligible in comparison to the spatial scale over which the electric field varies in particular in nanophotonic structures where large sub-wavelength field gradients are often present. Furthermore, while atoms are rotationally symmetric this is not the case for quantum dots where asymmetries, e.g., may arise from an inhomogeneous confinement potential that shifts electrons and holes differently due to their different effective masses. In the second subsection, we will review how the theory of QED can be extended beyond the dipole approximation and discuss the experimental observation of the breakdown of the dipole approximation.

\subsection{The phonon-mediated long-range Purcell effect}

\begin{figure}
\includegraphics[width=\columnwidth]{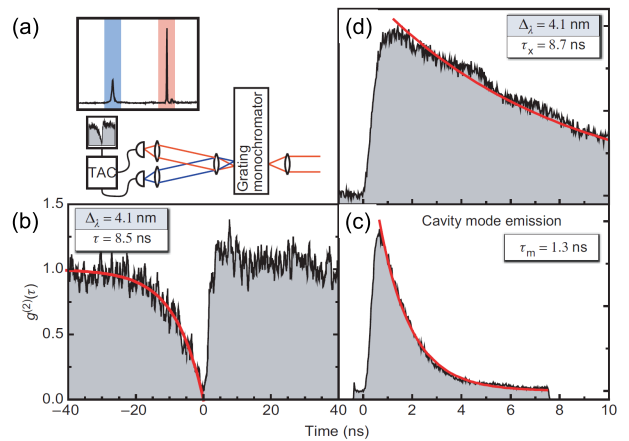}
\caption{\label{Figure_Hennessy} Observation of non-resonant coupling between a quantum dot and a photonic-crystal cavity. (a) Experimental setup and spectrum showing the cavity (blue) and quantum-dot (red) emission. (b) Cross-correlation between quantum-dot and cavity emission. (c) Decay curve of the quantum dot. (d) decay curve of the cavity-mode emission. Reprinted with permission from Ref.~\onlinecite{Hennessy2007}, copyright (2007) Macmillan Publishers Ltd.}
\end{figure}

Already the first cavity QED experiments on photonic-crystal cavities revealed that the simple textbook Jaynes-Cummings model~\cite{HarocheRaimondBook} that was developed for atoms in cavities did not explain the observations well. Hence it was found that a quantum dot could couple efficiently to the cavity mode even when the quantum dot was detuned many linewidths away from the cavity resonance. While such coupling was already present in the first experiments~\cite{Yoshie2004}, this surprising effect was clearly pinpointed in the work of Hennessy et al.~\cite{Hennessy2007} where cross-correlations between the cavity line and the quantum-dot line proved that the coupling could range as far as 4.1 nm, see Fig.~\ref{Figure_Hennessy}. The broadband coupling can be due to two effects: i) at a pump power far above saturation of the exciton ground state the quantum dot emits not on a single transition but a multitude of different levels show up due to a variety of possible charge configurations~\cite{Winger2009PRL}. ii) the coupling of the quantum dot to longitudinal acoustic phonons implies that the coupling range can be significantly enhanced since a quantum dot detuned to the red (blue) side of the cavity can emit a photon in the cavity if simultaneously a phonon is absorbed (emitted) to provide (carry away) the energy difference. Mechanism i) plays an essential role when the quantum dots are strongly pumped for instance in the case of quantum-dot lasers~\cite{Strauf2006}. In the context of cavity QED and single-photon emission mechanism ii) is most relevant, since multi-charge effects can be suppressed by controlling the excitation process. In experimental studies of the dynamics of a single quantum-dot line tuned through a cavity the sole influence of phonons can be probed  and broadband phonon-mediated Purcell enhancement has been reported that extends much further than predictions from the Jaynes-Cummings model~\onlinecite{Madsen2012}. The experimental data can be quantitatively understood by a full microscopic model of LA phonon depahsing of the quantum dot in the cavity. In the weak-coupling regime the quantum dot decay rate can be expressed as~\cite{Kaer2010PRL,Kaer2012PRB}
\be
\Gamma= \gamma+2g^2\frac{\gamma_{\text{tot}}}{\gamma^2_{\text{tot}}+\Delta^2}\left[ 1+\frac{1}{\hbar^2 \gamma_{\text{tot}}} \Phi(\Omega=\Delta)\right],
\label{gamma-phonons}
\ee
where $g$ is the light-matter coupling strength and $\gamma_\text{tot}=(\gamma+\kappa)/2$, where $\gamma$ is the decay rate associated with coupling to radiation modes and nonradiative recombination and $\kappa$ is the cavity decay rate. $\Phi(\Omega=\Delta)$ is the effective phonon density
experienced by the QD at the phonon frequency, $\Omega$, determined by the detuning, $\Delta$. This expression is an extension of the Jaynes-Cummings result where the interaction with phonons enters through the effective phonon density of states that is the responsible quantity describing all aspects of phonon-induced decoherence of the studied quantum dot. The relation constitutes an interesting link between mechanical degrees of freedom (i.e., phonons) and the radiative dynamics of the quantum dot. The exciton-phonon coupling is enhanced by the cavity (through the coupling strength $g$) but constitutes a different coupling mechanism than the photon-phonon interaction usually exploited in the field of quantum optomechanics, e.g., for cooling of mechanical objects~\cite{Kippenberg2008,Favero2009,Chan2011}. Such exciton-phonon coupling between a single quantum dot and a nanomembrane has been proposed as a path towards ground-state cooling of the membrane~\cite{Wilson-Rae2004}, and the first experiment in that direction demonstrated effective cooling using many carriers generated in bulk GaAs~\cite{Usami2012}. The relation of Eq. (\ref{gamma-phonons}) also offers a way of probing the energy dependence of the effective phonon density of states by comparing experiment to theory~\cite{Madsen2012}. Consequently, quantitative measurements of phonon decoherence processes can be extracted from cavity-QED experiments, which is essential since a thorough understanding of the complex phonon-dephasing behavior is required in order to generate highly indistinguishable photons for quantum-information processing~\cite{Kaer2012Preprint}.

\subsection{Breakdown of dipole approximation}
Quantum dots are spatially extended emitters that have inherently asymmetric wave functions. The most simple description of the carrier confinement in quantum dots restricts to only a two-band effective-mass model with strain that has proven to explain the dynamics of quantum dots in non-structured photonic media very well~\cite{Stobbe2009}. Even in such a simplistic model the electron and hole wave functions differ due to the difference of their effective masses. Asymmetries between electron and holes wave functions eventually imply that higher-order multipolar emission processes may alter the decay rate of dipole-allowed transitions. These effects can be enhanced geometrically when embedding the emitters in nanostructures with strongly varying electric fields.

The general theoretical framework for a description of spontaneous emission from two-level quantum dots beyond the dipole approximation has been put forward in Ref.~\onlinecite{Stobbe2012}. The radiative decay rate of an emitter of arbitrary size and shape embedded in any photonic structure is given by
\begin{align}
&\Gamma_\text{rad}(\mathbf{r}_0,\omega,\hat{\mathbf{e}}_\mathbf{d}) = \nonumber \\
& \frac{2 q^2}{\hbar m_0^2 c^2 \epsilon_0} \left| \mathbf{p}_\text{cv} \right|^2 \int \text{d}^3\mathbf{r} \int \text{d}^3\mathbf{r}'
\chi (\mathbf{r}_0,\mathbf{r},\mathbf{r})  \chi^\ast(\mathbf{r}_0,\mathbf{r}',\mathbf{r}')\nonumber
\\ &\times
 \left( \hat{\mathbf{e}}_\mathbf{d}^T \cdot \text{Im}\left\{\mathbf{G}( \mathbf{r},\mathbf{r}',\omega ) \right\} \cdot \hat{\mathbf{e}}_\mathbf{d} \right),\label{eq:NLIF_in_terms_of_G}
\end{align}
where $\mathbf{p}_\text{cv}$ is the Bloch matrix element describing the transition strength of the bulk crystal (related to the Kane Energy) and $\chi(\mathbf{r}_0,\mathbf{r}_\text{e},\mathbf{r}_\text{h})$ is envelope function of the exciton with electron and hole coordinates $\mathbf{r}_\text{e}$ and $\mathbf{r}_\text{h}$, respectively, which is centered at $\mathbf{r}_0$. This expression simplifies to the result valid in the dipole approximation, cf. Eqs.~(\ref{eq:OS1}), (\ref{eq:LDOS_in_terms_of_G}), and (\ref{eq:Gamma_rad}) in the limit that the Green's tensor, $\mathbf{G}(\mathbf{r},\mathbf{r}',\omega)$, varies insignificantly over the spatial extent of the emitter and in the strong-confinement regime for which $\chi(\mathbf{r}_0,\mathbf{r}_\text{e},\mathbf{r}_\text{h}) = F_\text{e}(\mathbf{r}_e) F_\text{h}(\mathbf{r})$. Interestingly, the light-matter degrees of freedom are found to be strongly intertwined in this theory as opposed to the case of the dipole approximation where the rate factorizes into a part related to the emitter (the oscillator strength) and a part related to the electromagnetic field (the LDOS), cf. Eq.~(\ref{eq:Gamma_rad}). This has the interesting consequence that the ability to change the radiative rate, i.e., to induce Purcell enhancement, is not determined solely by how well the emitter is positioned relative to the the local electric field maximum as it is the case for dipoles. Instead the Purcell enhancement is determined by a delicate and coherent interplay between the quantum dot wave function and the electromagnetic field. This leads to novel opportunities for controlling light-matter interaction by engineering in concert both electronic and photonic degrees of freedom.

\begin{figure}
\includegraphics[width=\columnwidth]{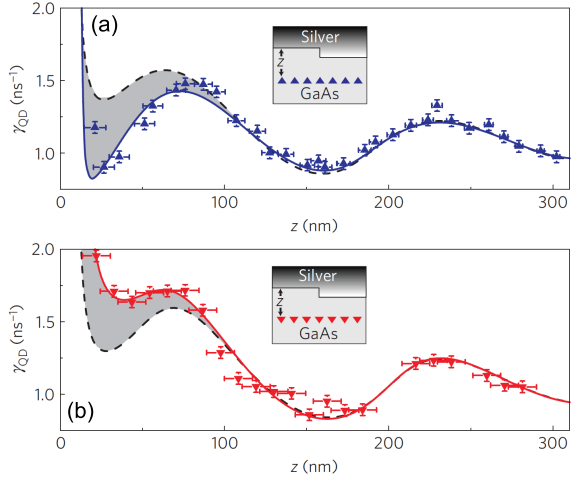}
\caption{\label{Figure_Mads} Experimental demonstration of light-matter interaction beyond the dipole approximation. (a) Measured spontaneous-emission rates (blue points) from an ensemble of quantum dots  placed at different distances, $z$, to a silver mirror. At distances below $100\:\text{nm}$ the measurements are systematically below the predictions from dipole theory that is described by the LDOS (dashed curve). (b) Similar measurements as in (a) but using an inverted structure, where the quantum dots are placed upside down relative to the mirror. Here the measurements (red points) are systematically above dipole theory. Both measurements are in good agreement with theory of spontaneous emission beyond the dipole approximation (solid blue and red lines). Reprinted with permission from Ref.~\onlinecite{Andersen2011}, copyright (2011) Macmillan Publishers Ltd.}
\end{figure}

The experimental demonstration of the breakdown of the dipole approximation was presented in Ref.~\onlinecite{Andersen2011}. In this experiment a simple nanostructure was chosen with a readily calculable yet strongly spatially varying local electric field, which was a silver metal mirror deposited on top of a GaAs substrate with quantum dots placed at different distances from the mirror. Figure~\ref{Figure_Mads} shows the measured decay rate of the quantum dots as a function of distance after depositing metal either on top (direct structure; Fig.~\ref{Figure_Mads}(a)) or on the bottom side (inverted structure; Fig.~\ref{Figure_Mads}(b)) of the substrate. These two different structures were fabricated in order to investigate the decay dynamics when inverting the quantum dots relative to the mirror. While a point-dipole source is invariant under such an inversion, the additional mesoscopic light-matter interaction terms are predicted to depend sensitively on this orientation. Indeed a theory obtained by Taylor expanding the light-matter interaction to first order beyond the dipole approximation was found to explain the experimental data in Fig.~\ref{Figure_Mads} well, where enhanced (suppressed) excitation of surface plasmon polaritons was observed for the inverted (direct) structure compared to the prediction from dipole theory. This experiment is a direct experimental demonstration of the prediction that both the position as well as the spatial extent and asymmetry of the quantum-dot wave function determine the decay rate of the quantum dot beyond the dipole approximation. Thus, it may provide the first step towards fully exploiting the opportunities of enhancing light-matter coupling by tailoring the quantum dot wave function together with the nanophotonic structure.

\section{Conclusions and future directions}
The physics of quantum dots embedded in photonic nanostructures is rich and fascinating. The research field has developed significantly in recent years with important progress both in theory and experiment, and thorough and quantitative understanding of the basic physical processes and how to control them is now at place. Based on this progress it appears realistic to start exploring more complex quantum systems where several quantum dots are coupled in nanophotonics networks with the long-term goal of establishing a platform for scalable quantum-information processing. In the present manuscript, we have reviewed the basic optical properties of InGaAs quantum dots including radiative and non-radiative decay processes. The detailed understanding of these processes allow using the quantum dots as sensitive local probes for complex nanophotonic structures. In the present review we have focused primarily on photonic crystals that can be employed for a range of different functionalities. In a photonic band gap, spontaneous emission can be suppressed and the state of the art is a 70-fold inhibition of the radiative rate of a single quantum dot. The experimental progress on QED with quantum dots in photonic-crystal cavities and waveguides was furthermore discussed including the relevant figures of merit, i.e., the Purcell factor and the $\beta$-factor. Both photonic-crystal cavities and waveguides have very promising figures of merits enabling strong coupling of light and matter and near-unity channeling of single photons to a propagating single mode, respectively. Finally we discussed that solid-state QED systems have unique properties distinguishing them from their atomic counterparts. In particular we saw that the coupling to phonons may give rise to broadband Purcell enhancement and that the dipole approximation for quantum dots is not always valid. Going beyond the dipole approximation may be employed for enhancing light-matter even more than possible for dipoles.

\begin{figure}[tb]
\includegraphics[width=83mm]{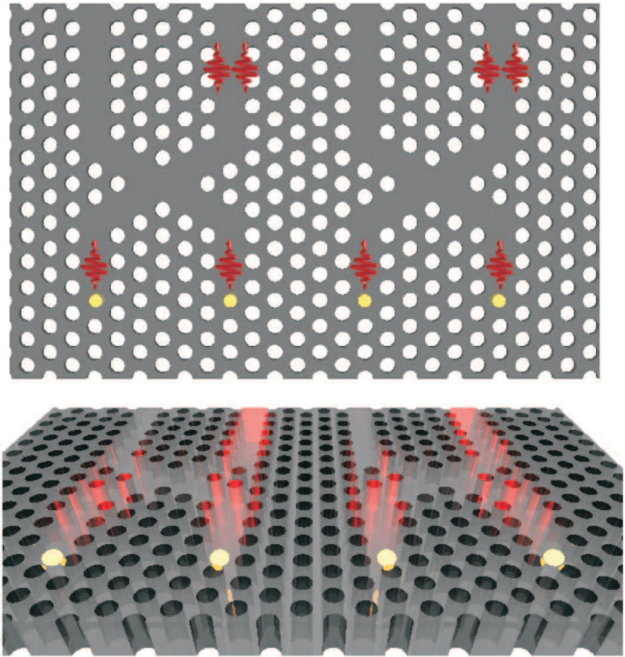}
\caption{Illustration of basic ingredients required for scalable quantum network based on quantum dots embedded in photonic-crystal waveguides. Four quantum dots (yellow spheres) are embedded in individual photonic-crystal waveguides as shown in top view (upper figure) and side view (lower figure). Each quantum dot emits single-photon wavepackets on demand with high efficiency to the waveguide and the photons subsequently interfere on integrated beam splitters enabling quantum processing.}
\label{Figure_scalability}
\end{figure}
The future research directions for quantum dots in photonic nanostructures are likely to be centered around scaling the simple functionalities implemented so far to larger quantum architectures integrated on an optical chip. The local light-matter coupling efficiency found in photonic crystals is unprecedented by other methods and constitutes a very promising starting point for such a research program. Figure \ref{Figure_scalability} illustrates a simple quantum network consisting of individual quantum dots efficiently coupled to photonic-crystal waveguides. The ability to electrically tune the quantum dots may enable interfering triggered single photons from the quantum dots on a chip, thus enabling deterministic quantum processing. Another attractive functionality would be to exploit the potentially giant nonlinearity of a quantum dot efficiently coupled to the photonic-crystal waveguide for novel quantum algorithms. The fundamental limits of the scalability set by environmental decoherence still remains to be developed and would depend as well on the particular quantum protocols being targeted.

We are very grateful to all the colleagues that have contributed to the work presented in this article: Jeppe Johansen, Philip Kristensen, Qin Wang, Toke Lund-Hansen, Brian Julsgaard, Henri Thyrrestrup, Kristian H{\o}eg Madsen, Per Kaer Nielsen, and Mads Lykke Andersen.
We would like to thank the Villum Foundation, the Danish Council for Independent Research (Natural Sciences and Technology and Production
Sciences) and the European Research Council (ERC consolidator grant "ALLQUANTUM") for financial support.

\end{document}